\documentclass{iaus}
\usepackage{graphicx}

\title[The Nature of ULXs from Optical/IR data] 
{On the Nature of Ultra-Luminous X-ray Sources from Optical/IR Measurements}

\author[Cropper \etal]   
{Mark Cropper$^1$, Chris Copperwheat$^1$, Roberto Soria$^{2,1}$, Kinwah Wu$^1$}

\affiliation{$^1$Mullard Space Science Laboratory, University College London, Holmbury St Mary, Dorking, Surrey RH5 6NT, UK \\[\affilskip]
$^2$Centre for Astrophysics, Harvard Smithsonian Astrophysical Observatory, 60 Garden St, Cambridge, MA 02138, USA}

\pubyear{2006}
\volume{238}  
\pagerange{001--999}
\date{??? and in revised form ???}
\jname{Black Holes: from Stars to Galaxies -- across the Range of Masses}
\editors{V. Karas \& G. Matt, eds.}
\begin{document}
\newcommand{\dg} {^{\circ}}
\outer\def\gtae {$\buildrel {\lower3pt\hbox{$>$}} \over
{\lower2pt\hbox{$\sim$}} $}
\outer\def\ltae {$\buildrel {\lower3pt\hbox{$<$}} \over
{\lower2pt\hbox{$\sim$}} $}
\newcommand{\ergscm} {ergs s$^{-1}$ cm$^{-2}$}
\newcommand{\ergss} {ergs s$^{-1}$}
\newcommand{\ergsd} {ergs s$^{-1}$ $d^{2}_{100}$}
\newcommand{\pcmsq} {cm$^{-2}$}
\newcommand{\ros} {{\it ROSAT}}
\newcommand{\xmm} {\mbox{{\it XMM-Newton}}}
\newcommand{\exo} {{\it EXOSAT}}
\newcommand{\sax} {{\it BeppoSAX}}
\newcommand{\chandra} {{\it Chandra}}
\newcommand{\hst} {{\it HST}}
\def\rchi{{${\chi}_{\nu}^{2}$}}
\newcommand{\Msun} {$M_{\odot}$}
\newcommand{\Rsun} {$R_{\odot}$}
\newcommand{\Mwd} {$M_{wd}$}
\newcommand{\Mbh}{$M_{\bullet}$}
\def\Mdot{\hbox{$\dot M$}}
\def\mdot{\hbox{$\dot m$}}
\def\mincir{\raise -2.truept\hbox{\rlap{\hbox{$\sim$}}\raise5.truept
\hbox{$<$}\ }}
\def\magcir{\raise -4.truept\hbox{\rlap{\hbox{$\sim$}}\raise5.truept
\hbox{$>$}\ }}

\maketitle

\begin{abstract}
We present a model for the prediction of the optical/infra-red emission from ULXs. In the model, ULXs are binary systems with accretion taking place through Roche lobe overflow. We show that irradiation effects and presence of an accretion disk significantly modify the optical/infrared flux compared to single stars, and also that the system orientation is important. We include additional constraints from the mass transfer rate to constrain the parameters of the donor star, and to a lesser extent the mass of the BH. We apply the model to fit photometric data for several ULX counterparts. We find that most donor stars are of spectral type B and are older and less massive than reported elsewhere, but that no late-type donors are admissable. The degeneracy of the acceptable parameter space will be significantly reduced with observations over a wider spectral range, and if time-resolved data become available.
\end{abstract}

\firstsection 
\section{Introduction}


Ultra-luminous X-ray sources (ULXs) are non-nuclear X-ray sources in nearby galaxies with inferred luminosity $>$few$\times10^{39}$ \ergss. This luminosity exceeds the Eddington luminosity of a 20\Msun\/ black hole (BH) (an observational overview is available in \cite{Fabbiano_04}). While these objects are generally agreed to be binary systems, the nature of their constituents is still controversial. Their emission could be as a result of sub-Eddington accretion rates onto intermediate mass black holes (IMBH) with masses $\sim200-1000$\Msun, (\cite{Colbert_99}), super-Eddington accretion onto stellar-mass BH (\cite{Begelmann_02}, \cite{King_01}) or Eddington accretion onto BH with masses in the range $\sim 50-200$ \Msun\/ (\cite{Soria_06}). 

Recently, reasonably secure optical counterparts for these systems have been identified, mostly using \hst\/ observations. This has opened a new channel of investigation into the nature of ULX. The optical/infrared emission is derived from the irradiated mass donor star and disk, so it is essential to model these appropriately in both the spectral and time domain if system parameters such as the mass and radius of the mass donor star and the mass of the BH are to be constrained. This paper describes such a model for the optical/infrared emission, and summarises some of the constraints that can be derived from its application to optical/infrared data. The objectives of this work are (a) to provide constraints on the possible optical counterparts of ULXs, eliminating those candidates which are inconsistent with the predicted colours/variability; (b) to determine the characteristics of the ULX constituent parts as accurately as possible; (c) to constrain the origin of ULXs and (d) to make predictions for future observations. More detailed expositions can be found in \cite{Copperwheat_05} and \cite{Copperwheat_06}.

\vspace*{-4mm}
\section{The model}\label{sec:next}

The compact object in the model is a BH of mass in the range $10-1000$\Msun. The mass donor star fills its Roche lobe, and accretion takes place through Roche lobe overflow into an accretion disk. We assume the disk to be a standard thin disk, tidally truncated at a radius 0.6 of that of the distance to the $L_1$ point. The mass donor star evolves according to the isolated star evolutionary tracks of \cite{Lejeune_01}. We ignore the effects of mass transfer on the star. We assume also that mass transfer is driven by the nuclear evolution of the mass donor. The model includes the Roche geometry, gravity and limb darkening, disk shadowing, radiation pressure according to the prescription of \cite{Phillips_02}, the evolution of the companion and system orientation effects (inclination and binary phase). The X-ray irradiation is assumed to be isotropic.

\begin{figure}
\center{
\includegraphics[width=0.85\columnwidth]{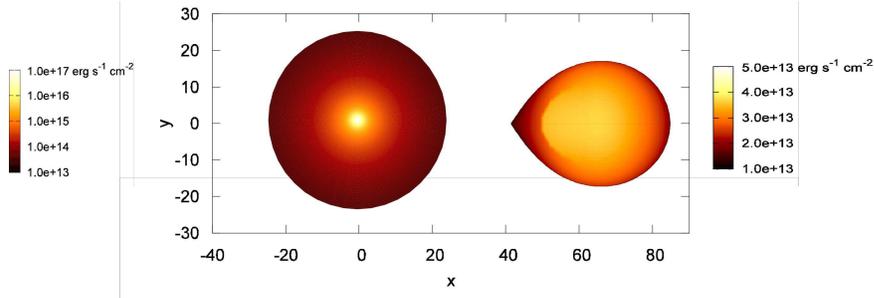}}
\caption{The variation in intensity $B(\tau)$ with $\tau=2/3$ for (right) an irradiated O5V star and (left) a disk for a BH mass of 150\Msun. The plot shows the view onto the orbital plane with the labelled distances in units of \Rsun. Note that the intensity scales are logarithmic for the disk and linear for the star. }
\label{fig:fig1}
\end{figure}

The irradiation of the disk and star is handled according to a formulation by \cite{Wu_01} which is based originally on the grey stellar irradiation model of \cite{Milne_26} and incorporates the different opacities of the irradiated surface to hard and soft X-rays (the X-ray hardness ratio is an input parameter). The effective temperature of the irradiated star or irrdiated disk is a superposition of the irradiated and natural temperatures {\it i.e.}
$T_{eff} = \left( \frac{\pi}{\sigma} B_{x}(2/3) + T^{4}_{unirr}\right)^{1/4}$ where 
$B_x(\tau)$ is derived in \cite{Copperwheat_05}. Figure~\ref{fig:fig1} provides an example of the irradiated disk and star.

\begin{figure}
\includegraphics[width=\columnwidth]{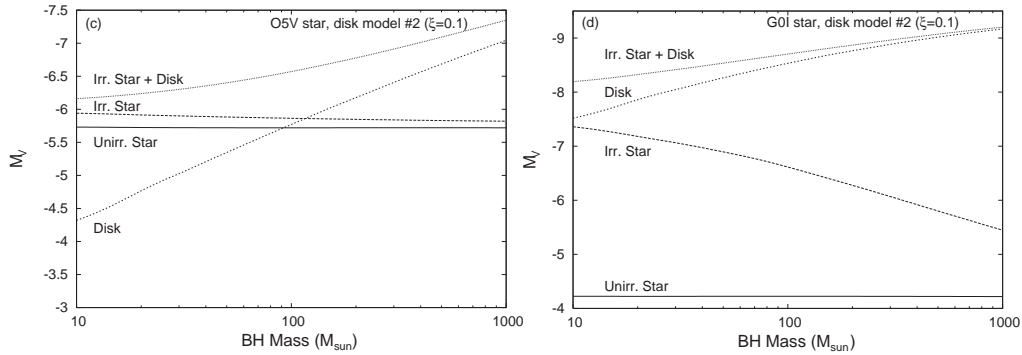}
\caption{The $V$ band absolute magnitudes for an un-irradiated and irradiated O5V star and accretion disk (left) and for a G0I supergiant with disk (right), shown as a function of BH mass. Here $L_x = 10^{40}$ \ergss, $\cos{i} = 0.5$ and the star is at superior conjunction.}
\label{fig:fig2}
\end{figure}

Using this model we can predict the different contributions of the constituents of the ULX as a function of (for example) BH mass and donor star spectral type, as shown in Figure~\ref{fig:fig2}. For high-mass BHs, the disk is large, and hence disk emission dominates. In addition, irradiation effects are much larger for late-type stars. We make predictions of the optical/infrared flux. We find that there is generally better system parameter discrimination at infrared wavelengths. We also predict the variability timescale.  Because of the axial symmetry of our disks, these are dominated by the (modified) ellipsoidal variations from the Roche-lobe-filling donor stars. The variability timescale is typically days, dependent on BH mass.

The photometric predictions can be compared to observations of ULX counterparts in different wavebands, and acceptable model parameter regimes determined from $\chi^2$ fitting to the observations. If only single-epoch photometric observations are available in 2 or 3 bands the model fits are underconstrained, leading to degenerate solutions. An example is shown in Figure~\ref{fig:fig3}(left). Nevertheless, even with limited photometric data much of the possible parameter space can be eliminated, and with additional wavelength coverage particularly in the infrared, degeneracies can be reduced further. 

We add a further constraint for the $\chi^2$ fitting using the mass transfer rate, which is determined by the evolution of the stellar radius (from the evolutionary tracks) compared with the evolution of the Roche lobe radius (\cite{Wu_97}, \cite{Ritter_88}). We measure the mass transfer rate from the X-ray luminosity assuming an accretion efficiency $\eta=0.1$ appropriate for a BH, and select only those secondary stars which are evolving in radius on nuclear timescales in such a way as to provide the measured mass transfer rate.

\vspace*{-4mm}
\section{Fits to data}

We have gathered \hst\/ and VLT photometric data available up to mid-2006 on the most luminous ULX counterparts, where we can be reasonably certain that mass transfer is driven by Roche lobe overflow. The input data for M81 X-6, NGC 4559 X-7, M101 ULX-1, NGC 5408 ULX, Holmberg II ULX, NGC 1313 X-2 (C1) and NGC5204 ULX are given in table 1 of \cite{Copperwheat_06} and for M51 X5/9 in \cite{Copperwheat_07}. We fitted our model to these photometric data, using the additional mass transfer constraint from X-ray measurements. 

\begin{figure}
\center
\includegraphics[angle=270,width=0.92\columnwidth]{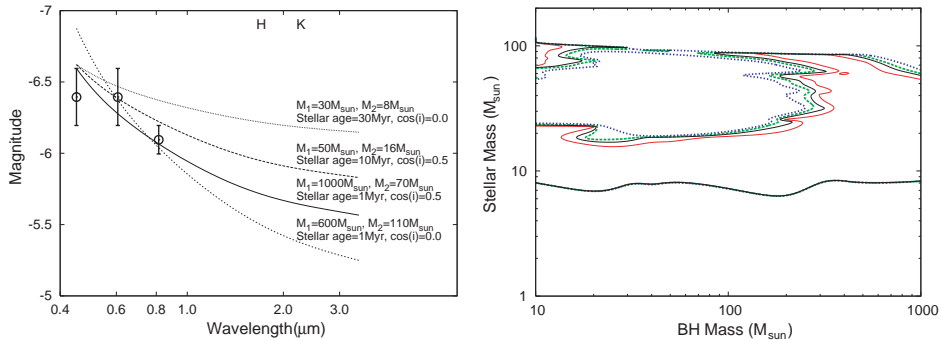}
\caption{(left) Absolute magnitude as a function of wavelength for different combinations of BH (M$_1$) and donor star (M$_2$) and inclination {\it i} compared to photometric measurements for NGC5408 ULX, indicating the degeneracy of possible solutions when no mass transfer constraints are applied. (right) Acceptable parameter space for the mass donor star projected onto the donor mass/BH mass plane when the mass transfer constraint is included. Contours are at 68, 90 (solid line), 95 and 99\% confidence levels.}
\label{fig:fig3}
\end{figure}

Figure~\ref{fig:fig3} (right) shows an example of the allowed parameter space in the donor star mass vs BH mass plane from the $\chi^2$ fitting. We also produced similar plots for the donor star radius and donor star age, which we determined from the mass-radius relation from the evolutionary tracks. More details can be found in \cite{Copperwheat_06}.

These fits provide the current spectral type of the mass donor star. By tracing to earlier times along the evolutionary tracks, the ZAMS spectral type can be predicted assuming the mass loss has not significantly altered the evolution.

\vspace*{-4mm}
\section{Main outcomes}

The optical/infrared emission from our binary ULX model is significantly different from models assuming unirradiated companion stars and no disks, hence it is not adequate to assume standard colours from single stars to determine the mass donor characteristics in ULXs.

The model fits to the currently available data do not provide strong constraints on the BH mass, mainly because of the unknown orientation of the system. Depending on inclination, upper or lower limits can, however, be set.  For example, the BH mass in NGC 5204 is $<240$ \Msun\/ for $\cos{i}=0.5$, while that in NGC 1313 X-1 is $<100$ \Msun. The reason is that the signature of a disk can be almost entirely hidden for $\cos{i}=0.0$, in which case little information can be obtained on the size of the primary Roche lobe.

The mass, radius and age of the donor star are, however, more tightly constrained. Typical ages range from $10^7-10^8$ years, with typical ZAMS masses $5-10$ \Msun, with some up to 50 \Msun. In general we find that the mass donor stars are less massive and older than generally quoted in the literature from less comprehensive modeling -- this is to be expected given the effects of irradiation. We find that none of the systems are found to contain late-type mass donor stars, whatever disk component is admitted.

The preference for donor stars of spectral type B is interesting. The high mass transfer rates and modest donor star masses require that ULX lifetimes are short (this is true also for higher mass donors which in any case have short lifetimes). The duration that the donor star has been in contact with its Roche lobe is an important parameter for ULXs. If it has been in contact for some Myr, especially in the case of B-type ZAMS stars, then binary evolution models will be required.

The diagnostic capability of our model for the BH mass and donor star characteristics will improve significantly as more filter bands, and particularly time-resolved data become available.

\begin{acknowledgments}
R. Soria acknowledges support from a Marie Curie Fellowship from the EC.
\end{acknowledgments}

\end{document}